\documentclass[letterpaper,twocolumn,10pt]{article}
\PassOptionsToPackage{hypertexnames=false}{hyperref}
\usepackage{usenix}

\usepackage[color,reflinks]{crypto}
\IfFileExists{bbm.sty}{\usepackage{bbm}}{}
\providecommand{\mathbbm}[1]{\mathbf{#1}}
\usepackage{enumitem, array, bm, multirow}
\usepackage{tabularx}
\usepackage{amsmath,amsfonts,amsthm}
\newtheorem{theorem}{Theorem}[section] 
\newtheorem{definition}[theorem]{Definition} 
\newtheorem{lemma}{Lemma}

\usepackage{graphicx}
\usepackage{xcolor}
\usepackage{diagbox}
\IfFileExists{siunitx.sty}{\usepackage{siunitx}}{}
\usepackage{makecell}
\usepackage{pifont}
\IfFileExists{algpseudocode.sty}{%
  \usepackage{algpseudocode}
  \let\AlgoFor\For
  \let\AlgoEndFor\EndFor
  \let\AlgoIf\If
  \let\AlgoElse\Else
  \let\AlgoEndIf\EndIf
  \let\AlgoComment\Comment
  \let\AlgoReturn\Return
}{}
\usepackage{booktabs}
\usepackage{cleveref}
\usepackage{amssymb}
\usepackage{changes}
\IfFileExists{algorithm2e.sty}{%
  \usepackage[ruled,vlined]{algorithm2e}
  \SetKwComment{CommentTri}{$\triangleright$~}{}
}{%
  \usepackage{float}
  \floatstyle{ruled}
  \newfloat{algorithm}{tbp}{loa}
  \floatname{algorithm}{Algorithm}

  \providecommand{\Return}{\textbf{return} }
  \providecommand{\tcp}{}
  \RenewDocumentCommand{\tcp}{s O{} m}{\hfill\(\triangleright\)~##3\par}
  \providecommand{\SetKwComment}[3]{}
}
\IfFileExists{algpseudocode.sty}{%
  \let\For\AlgoFor
  \let\EndFor\AlgoEndFor
  \let\If\AlgoIf
  \let\Else\AlgoElse
  \let\EndIf\AlgoEndIf
  \let\Comment\AlgoComment
  \let\Return\AlgoReturn
}{}

\IfFileExists{float.sty}{\usepackage{float}}{}
\floatstyle{ruled}
\newfloat{functionality}{tbp}{lofunc}
\floatname{functionality}{Functionality}
\crefname{functionality}{Functionality}{Functionalities}
\Crefname{functionality}{Functionality}{Functionalities}

\newcommand{\secret}[1]{{ \langle #1 \rangle}}

\newcommand{\gauss}{\mathbf{N}}
\newcommand{\lap}{\mathsf{Lap}}
\newcommand{\round}{\mathrm{round}}
\newcommand{\bern}{\mathsf{Bern}}

\definecolor{tododarkblue}{RGB}{0,45,114}
\definecolor{todolightblue}{RGB}{74,144,226}

\newcommand{\mypara}[1]{\vspace*{0.05in}\noindent\textbf{#1.} \xspace}

\begin{document}

\date{}

\title{\Large \bf Revisiting Continuous Noise Sampling for Multi-Party Differential Privacy}

\author{
Yucheng Fu, Tianhao Wang\\
University of Virginia\\
\{zdp8uu, tianhao\}@virginia.edu
}
\maketitle

\begin{abstract}

Combining secure multi-party computation (MPC) with differential privacy (DP) enables multiple parties to release aggregate statistics without a trusted curator, and the core primitive is the protocol to sample noise from a continuous distribution under finite-precision arithmetic. In this paper, we revisit the continuous noise sampling protocols and present several improvements in both security and efficiency.

We start by identifying a vulnerability in widely used sample-and-scale constructions. We demonstrate that the scaling operation in arithmetic circuits confines the noise to a sparse, publicly known set of values, so that an adversary can observe the released noisy queries and decide which dataset produced them. As concrete demonstrations, we instantiate attacks on two systems employing such ``flawed'' sampling protocols: Orchard (OSDI'20) for DP secure aggregation and DP-BREM$^+$ (USENIX Sec'25) for DP federated learning. We report a near-$100\%$ attack success rate on both systems, under any noise scaler $s\geq 2$ used in practice.

The leakage we reveal is intrinsic to the scaling operation, and direct repairs either substantially sacrifice utility or add significant precision bits to make the sampling more expensive. 
To address the security and efficiency issues together, we turn to discrete sampling at the granularity of individual biased bits. We make several optimizations to the sampler and prove its security. Our implementation achieves $4\times \sim 612\times$ speedup over existing secure discrete samplers and orders-of-magnitude speedup over the insecure sample-and-scale paradigm, with negligible utility loss compared to the ideal continuous mechanism.

\end{abstract}

\section{Introduction}

Differential Privacy (DP) has been recognized as the \emph{de facto} notion for protecting individuals' privacy in the output of algorithms. It has been widely employed in both industry and academia (e.g., Apple~\cite{team2017learning, pease2016engineering}, Google~\cite{bittau2017prochlo, erlingsson2014rappor}, Microsoft~\cite{ding2017collecting}, and LinkedIn~\cite{rogers2020differentially, rogers2020linkedin}). A common way to implement DP is the {\it centralized model}, which adds random noise to the algorithm's output before releasing it to the public. For example, a company can add random noise to the number of users who clicked an advertisement~\cite{reznichenko2014private, sun2022practical} or add noise to the gradient of its machine learning model to avoid training data being inferred~\cite{abadi2016deep, agarwal2021skellam, yu2021differentially}. 
However, the centralized model assumes a trusted server that collects all user data and executes the aggregation algorithm.

To implement DP without trusted servers, a line of work studies the {\it local model}, where each user randomizes their data point before uploading it to the server~\cite{cormode2018privacy, erlingsson2014rappor}. While removing the assumption of trusted servers, this model causes unavoidable utility degradation. Later, the {\it shuffled model} was proposed to mitigate utility loss in local models, but it still incurs higher error than the centralized model~\cite{cheu2019distributed}. To achieve the same utility as the centralized model, a promising approach is to leverage Secure Multi-party Computation (MPC)~\cite{beimel2008distributed, dwork2006our, mironov2009computational}. MPC is a cryptographic technique that allows two or more parties to jointly compute a function without knowing each other's inputs. Thus, one can design an MPC protocol among users and servers that simulates the DP algorithm in the centralized model while revealing only the noisy output. For example, starting with iOS 17, Apple leverages the Prio secure aggregation protocol~\cite{corrigan2017prio} alongside DP to privately collect photo metadata (e.g., tags and locations) for features like Memories and Places~\cite{appleml2024scenes}.
In such a DP algorithm with MPC, a challenge is designing a protocol that {\it securely samples random noise}. 

Ideally, the random noise used to achieve DP is drawn from a Laplace or Gaussian distribution, both defined over the real number domain $\mathbb{R}$. Even in the centralized model of DP, sampling noise from these continuous distributions requires extremely careful design because it is impossible for finite computers to represent numbers in $\mathbb{R}$, and the noise sampling in standard libraries of C++, Python, etc. requires evaluating transcendental functions (e.g., the natural logarithm for Laplace noise~\cite{knuth2007computer, press2007numerical} and the Marsaglia-polar / Box-Muller transformation for Gaussian noise~\cite{marsaglia1964convenient, lee2006hardware}). Many works have shown that generating noise by naively evaluating transcendental functions on floating-point representation may lead to failures of DP~\cite{jin2022we, mironov2012significance, holohan2024securing, ilvento2020implementing}, i.e., an adversary observing the output can distinguish the neighboring input datasets with a probability much higher than the DP guarantee allows.

When sampling noise in MPC, the risk of violating the DP guarantee can be even more severe. If the protocols are implemented in floating-point arithmetic MPC, all vulnerabilities in the centralized model are inherent. Moreover, instead of using floating-point representation, the MPC frameworks tend to use {\it fixed-point} representation~\cite{keller2020mp, knott2021crypten, mohassel2018aby3, demmler2015aby} in the arithmetic circuit. This is because the floating-point implementation incurs several-fold communication overhead compared to fixed-point ones~\cite{rathee2022secfloat}, which raises efficiency concerns in practice. Correspondingly, the noise sampling protocols also work under the arithmetic circuit with fixed-point representation and can be summarized as two categories: 

\begin{itemize}[leftmargin=*]
    \item {\it Distributed noise generation~\cite{ruan2023private, roy2020crypte,dwork2006our,goryczka2015comprehensive,bohler2021secure}.} The distributed noise generation protocols require each party to locally sample its partial noise and securely aggregate them in MPC. Although efficient, these methods compromise privacy, because an adversary corrupting $t$-out-of-$N$ parties can simply remove $t$ partial noise samples from the released DP query and weaken the privacy guarantee.
    \item {\it Sample-and-scale~\cite{gu2025dp,eigner2014differentially,pentyalacaps,pentyala2022training,roth2019honeycrisp,roth2020orchard}.} Another construction without privacy loss emulates the sampling algorithms of centralized DP in MPC, which evaluate an arithmetic circuit of approximated transcendental functions to draw the secret random number $\secret{r}$ from the standard Laplace / Gaussian distribution. Then $\secret{r}$ is multiplied by a public scaler $s$ such that $\secret{s\cdot r}$ comes from a distribution with variance required by the DP mechanism. Such arithmetic circuits can be evaluated by various generic MPC protocols with different security assumptions.
\end{itemize}
For the first category of continuous noise sampling, the trade-off between privacy and utility has been well-studied by existing works~\cite{dwork2006our, fu2024benchmarking, canonne2020discrete, bohler2021secure}, which conclude that for a $t$-out-of-$N$ corruption, we must either suffer a privacy loss by a factor of $1-t/N$, or enlarge the partial noise and thus suffer a utility loss by the same factor. 
In this paper, we revisit the second category, {\it sample-and-scale} which seems to avoid the privacy-utility trade-off. We observe that the status quo raises the following two challenges.

\begin{itemize}[leftmargin=*]
    \item {\it Challenge 1: unknown vulnerabilities.} The sample-and-scale constructions' DP guarantees are proven in the real-number field, where every operation runs at infinite precision. A real arithmetic circuit obviously cannot honor this assumption. It therefore remains unknown whether the released $\secret{s\cdot r}$ is still a legal noise sample, i.e., a value drawn from the target Laplace / Gaussian distribution and then rounded to the nearest representable number. Only in this case can we claim that the DP guarantee of sample-and-scale carries over without loss. 
    \item {\it Challenge 2: inefficiency.} Emulating the centralized sampler in MPC requires evaluating an approximated transcendental function with the arithmetic circuit, e.g., the logarithm for Laplace noise or the Box-Muller transform for Gaussian noise, which incur significant overhead in MPC. In our early-stage evaluation, even if we put the vulnerability aside, drawing a single noise value with the sample-and-scale method takes about ${19}$ seconds for Laplace and ${61}$ seconds for Gaussian, at scale $s=10$ and with $l=16$ bits in the fractional part\footnote{These results are measured under a two-party arithmetic secret-sharing protocol implemented with MP-SPDZ~\cite{keller2020mp} in a WAN setting, where the bandwidth is up to $1$~Gbps and the delay is up to $100$~ms.}. This result is far from practical. We therefore ask whether continuous noise sampling can be made practically efficient.
\end{itemize}

\subsection{Results and Contributions}

\begin{table*}[t]
\centering
\footnotesize
\begin{tabular*}{0.85\textwidth}{@{\extracolsep{\fill}} l c c c c @{}}
\toprule
\textbf{Paper} & \textbf{Venue} & \textbf{Transcendental Function} & \textbf{Computation for Scaling} & \textbf{Supported Query} \\
\midrule
Honeycrisp~\cite{roth2019honeycrisp}        & SOSP'19        & Unspecified                & Fixed-point multiplication    & Count \\
Orchard~\cite{roth2020orchard}              & OSDI'20        & Unspecified                & Fixed-point multiplication    & 17 SQL queries \\
PrivaDA~\cite{eigner2014differentially}      & ACSAC'14       & Inverse CDF of Laplace     & Floating-point multiplication & Sampling only \\
Pentyala et al.~\cite{pentyala2022training}  & Arxiv'22       & Box-Muller transformation  & Fixed-point multiplication    & Gradients \\
CaPS~\cite{pentyalacaps}                     & ICML'24        & Irwin-Hall approximation   & Fixed-point multiplication    & Histogram \\
DP-BREM$^+$~\cite{gu2025dp}                  & USENIX Sec'25  & Box-Muller transformation  & Fixed-point multiplication    & Gradients \\
\bottomrule
\end{tabular*}
\caption{Existing sample-and-scale protocols vulnerable to our attack. All of these sampling protocols first draw a standard noise sample with a transcendental function and scale it to provide the targeted DP guarantee. We note that PrivaDA~\cite{eigner2014differentially} is a special construction that multiplies the scaler in floating-point arithmetic, and we empirically show that it is also vulnerable to our attack in Section~\ref{sec:practical_attack}.}
\label{tab:vulnerable}
\end{table*}

\mypara{Uncovering vulnerability in sample-and-scale} We revisit the sample-and-scale procedure in the arithmetic circuit and uncover a vulnerability that arises for every scaler $s>1$. This range is exactly the one considered in practice. For example, to protect a single query with sensitivity $1$, a scaler $s>1$ corresponds to a privacy budget $\epsilon<1$ for the Laplace mechanism and $\epsilon < 5$ for the Gaussian mechanism.

\begin{itemize}[leftmargin=*]
    \item {\it Analysis (Section~\ref{sec:attack-uniform}).} We show that given two neighboring datasets, an adversary can distinguish which one produced the noised query by only observing the protocol's output, with a probability beyond what the DP guarantee allows. We confirm that this distinguishing probability increases dramatically with $s$ (in which case it should decrease in a correctly implemented DP mechanism), and reaches almost $100\%$ once $s \geq 2$, which completely breaks the DP guarantee. Moreover, the distinguishability remains even though the queries on these two datasets differ in a single bit, which is stronger than previous attacks that need the two queries to differ substantially or allow the adversary to pick special neighboring pairs~\cite{mironov2012significance,jin2022we,ilvento2020implementing,holohan2024securing}.
    \item {\it Attacks on existing systems (Section~\ref{sec:attack-experiments}).} For demonstration, we utilize the above vulnerability to launch membership inference attacks on two systems deploying the sample-and-scale sampling protocols: Orchard~\cite{roth2020orchard} for secure DP $k$-means and DP-BREM$^{+}$~\cite{gu2025dp} for DP federated learning. We report a near-$100\%$ success rate under the widely used privacy budget configurations. 
\end{itemize}
Our attack can be generalized to other sampling protocols that follow the sample-and-scale paradigm, rather than being specific to \cite{roth2020orchard,gu2025dp}. We summarize the protocols vulnerable to our attack in Table~\ref{tab:vulnerable}.

\mypara{Secure and efficient continuous noise sampling} To address the vulnerability, we design a new sampling protocol for continuous noise that comes with provable security and a concrete efficiency improvement over prior (in-)secure samplers.

\begin{itemize}[leftmargin=*]
    \item {\it Security (Section~\ref{sec:securing}).} We formalize the ideal functionality that exactly draws a noise sample and rounds it into the finite-precision fixed-point representation. We then extend previous sampling protocols for integer noise~\cite{fu2024benchmarking,dwork2006our,wei2023securely} to the fixed-point grid, and prove that it securely realizes the ideal functionality and closes the vulnerability we uncover.
    \item {\it Efficiency and Exactness (Section~\ref{sec:secure-experiments}).} We implement the secure sampling protocol and make several improvements on efficiency. As a result, our protocol achieves $2\times \sim 706\times$ speedup compared to the directly adapted secure implementations, and $190\times \sim 1{,}317\times$ speedup compared to the insecure sample-and-scale construction under the WAN setup. Moreover, its output distribution matches the one defined by the ideal functionality and achieves utility comparable to the ideal continuous mechanism. Our implementation is available at \url{https://github.com/yuchengxj/ContinuousNoise-Revisit}.
\end{itemize}

\subsection{Technical Overview}

\begin{figure}[t]
\centering
\includegraphics[width=0.9\columnwidth]{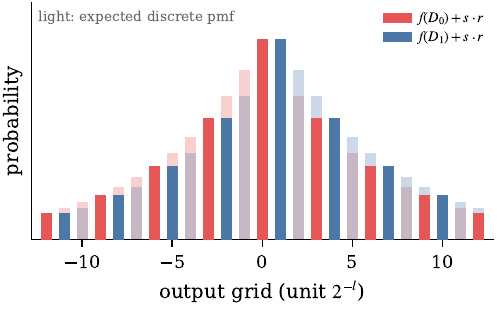}
\caption{Sparse and disjoint outputs of sample-and-scale on the fixed-point grid with interval $[-12, 12]\times 2^{-l}$. A correct sampler places mass on every grid point, shown as the light bars. Scaling the standard sample $r$ by $s$ instead reaches only one out of every $s$ grid points, shown as the solid bars, so that $f(D_0)+s\cdot r$ and $f(D_1)+s\cdot r$ never overlap.}
\label{fig:intro-sparsity}
\end{figure}

\mypara{Output sparsity of scaling in arithmetic circuit} Consider a fixed-point representation with $n$ bits in the integer part and $l$ bits in the fractional part. The sample-and-scale paradigm draws a standard sample $r$ from the standard Laplace $\lap(1)$ or Gaussian $\gauss(1)$, which is rounded onto the fixed-point grid with step size $2^{-l}$. Then, the protocol multiplies it by a public scaler $s$ to reach the target scale. Because $r$ already sits on the grid, if $s>1$, the scaled value $s\cdot r$ can only land on a sparser subset of grid points and never outputs the remaining values. As a result, consider a pair of distinct queries $f(D_0)$ and $f(D_1)$ produced by neighbouring datasets $D_0$ and $D_1$, the noised queries $f(D_0)+s\cdot r$ and $f(D_1)+s\cdot r$ will sit in two disjoint output sets, as shown in Figure~\ref{fig:intro-sparsity}. Hence, an adversary who knows the public scaler $s$ can read them off the output and tell the two neighboring datasets apart. 

There are two direct fixes to the above vulnerability. However, both of them come at a price. The first solution is rounding the query $f(D)$ onto the same sparse grid as the noise, which realigns the two grids but discards resolution and loses utility that grows with $s$. The second adds $\lceil\log_2 s\rceil$ guard bits to the sample so that $s\cdot r$ fills every grid point again. This one keeps the utility, yet it widens every value in the circuit by a factor of $(l+\lceil\log_2 s\rceil)/l$ and makes the already expensive sampler even slower. Moreover, both fixes require the precise evaluation of expensive transcendental functions, which motivates a sampler that is correct on the grid by construction.

\mypara{Adapting and parallelizing secure coin flipping} Our starting point is a simple observation. A fixed-point number is an integer multiple of the grid step $2^{-l}$, so sampling noise on the grid is the same as sampling an integer and reading it at resolution $2^{-l}$. Instead of drawing a standard sample and scaling it, we draw the integer directly from a discrete Laplace or discrete Gaussian with integer scale $t=s\cdot 2^{l}$, and then attach the factor $2^{-l}$. This sampler puts mass on every grid point, so the sparsity above disappears. Its output differs from the ideal continuous mechanism by only $O(2^{-l})$, which is negligible at the precisions used in practice.

Existing discrete samplers~\cite{fu2024benchmarking,dwork2006differential,wei2023securely,champion2019securely} heavily rely on a protocol that securely tosses biased coins (i.e., samples secret bits from a Bernoulli distribution). In fact, all these biased coins are independent from each other~\cite{dwork2006differential} and can be tossed in parallel. However, existing secret-sharing implementation~\cite{fu2024benchmarking} still tosses these coins sequentially and leads to a large number of communication rounds, which dominate the running time under a WAN. 
In our implementation, we parallelize all the coin flipping to achieve a number of communication rounds nearly independent from the number of samples and the number of bits in each sample (this is crucial in our setup with $l$ additional bits for the fractional part). In our evaluation, we observe a $476\times \sim 1{,}733\times$ reduction in the communication rounds, which is where most of our speedup over the directly adapted secret sharing sampler comes from. Under the two-party setup, we also surpass the constant round garbled circuit instantiation because of the lower total communication by factors of $4.5\times \sim 6.1\times$. 

% \mypara{Roadmap} The rest of this paper is organized as follows. In Section~\ref{sec:preliminary}, we introduce the preliminaries. In Section~\ref{sec:leak}, we present the two attacks on existing continuous noise generation protocols in MPC. In Section~\ref{sec:securing}, we describe a general method to secure the continuous noise generation protocols and evaluate the correctness and efficiency of this method. We discuss future works in Section~\ref{sec:discuss} and conclude in Section~\ref{sec:conclude}.

\section{Preliminaries}\label{sec:preliminary}
\subsection{Notations} 
We use $\lambda$ to denote the security parameter. In this paper, we focus on fixed-point representation. We write $\mathbb{F}_{n, l}$ for the set of fixed-point numbers with $n$ bits in the integer part and $l$ bits in the fractional part. In the arithmetic circuit, every $x\in\mathbb{F}_{n, l}$ is stored through its integer encoding $X=2^{l}x$, where the capital letter $X$ is a signed $(n+l)$-bit integer. In other words, the representable values are the integer multiples of $2^{-l}$ whose encoding satisfies $|X|<2^{n+l-1}$, so they are evenly spaced by $2^{-l}$, which we refer to as the fixed-point grid. 

\subsection{Secure Multi-party Computation}

Secure multi-party computation (MPC)~\cite{goldreich2001foundations, yao1982protocols} enables $N$ parties $P_1, \ldots, P_N$ to collaboratively evaluate a function $y = f(x_1, \ldots, x_N)$ on their private inputs $x_1, \ldots, x_N$, where party $P_i$ holds input $x_i$. The security guarantee ensures that each party learns only the output $y$ and nothing beyond what can be inferred from $y$ and its own input. In particular, no party can learn other parties' inputs or any intermediate values in the protocol execution.

A direct way to realize MPC for a given function $f$ is to represent $f$ as a circuit and execute it with a general-purpose protocol, using either Boolean circuits with bit-level operations (e.g., XOR, AND) or arithmetic circuits with field operations. In this paper, we focus on arithmetic circuits over integers and fixed-point numbers, which are typically implemented by secret sharing. The adversary is semi-honest when it follows the protocol while staying curious about privacy, and malicious when it may deviate from the execution.

\mypara{Secret Sharing}
In secret sharing-based protocols, each party ``shares'' its input $x$ to $N$ pieces $\{\secret{x}_1, \dots, \secret{x}_N\}$ (corresponding to $N$ parties) and sends them to the rest of the parties. For simplicity, we use $\secret{x}=\{\secret{x}_1, \dots, \secret{x}_N\}$ to denote the shares of all $N$ parties. Then, the parties jointly run a protocol that outputs $\secret{y}_i$ to each party $i$, which are then combined to plain output $y$ via some $y ={\sf Reveal}(\secret{y})$ protocol.

\subsection{Differential Privacy}

Differential privacy (DP), introduced by Dwork et al.~\cite{dwork2006differential},
is a standard notion of privacy for computation over datasets. Before defining DP, we first introduce the concept of neighboring datasets.

\begin{definition}[Neighboring Datasets]
Two datasets $D_0, D_1\in \mathbb{D}^*$ are said to be neighboring if they differ in exactly one record, denoted by $|D_0 \triangle D_1| =1$, where differing means both removal ($D_1$ is obtained by adding/removing any one record from $D_0$) and replacement ($D_1$ is obtained by replacing any one record from $D_0$).
\end{definition}

The sensitivity of a query function $f: \mathbb{D}^* \rightarrow \mathbb{R}$ measures the maximum change in the output due to a single record modification, defined as $\Delta_f = \max_{D_0,D_1} |f(D_0) - f(D_1)|$ where $|D_0 \triangle D_1| =1$.

\begin{definition}[Differential Privacy]\label{def:dp}
    Let $\epsilon>0$ and $\delta \in [0,1]$. A randomized algorithm $M: \mathbb{D}^* \rightarrow \mathbb{O}$ is said to be $(\epsilon,\delta)$-differentially private if for all neighboring datasets $D_0, D_1 \in \mathbb{D}$ and all possible outputs $\mathbb{S} \subseteq \mathbb{O}$:
    \[
        \Pr[M(D_0) \in \mathbb{S} ] \leq e^{\epsilon} \cdot \Pr[M(D_1) \in \mathbb{S} ] + \delta.
    \]
\end{definition}

\mypara{Laplace Mechanism} For a query $f$ with sensitivity $\Delta_f$, the Laplace mechanism~\cite{dwork2014algorithmic} achieves $\epsilon$-DP by
computing $f(D)+r$, where $r$ is drawn from the Laplace distribution with scale parameter $b = \frac{\Delta_f}{\epsilon}$, denoted by $\lap(b)$. The Laplace distribution has PDF $p(x;b) = \frac{1}{2b}\exp(-\frac{|x|}{b})$. 

\mypara{Gaussian Mechanism} For a query $f$ with sensitivity $\Delta_f$, the Gaussian mechanism~\cite{dwork2014algorithmic} achieves $(\epsilon, \delta)$-DP by computing $f(D)+r$, where $r$ is drawn from the Gaussian distribution with variance $\sigma^2=\frac{2\ln(1.25/\delta)}{\epsilon^2}$, denoted by $\gauss(\sigma^2)$. The Gaussian distribution has PDF $p(x;\sigma) = \frac{1}{\sqrt{2\pi}\sigma}\exp(-\frac{x^2}{2\sigma^2})$. There are recent improved results~\cite{balle2018improving,mironov2017renyi} about the Gaussian mechanism.  In this paper, for simplicity, we use the classic result.  

% We note that the above definitions and mechanisms of DP are mathematical abstractions assuming that the algorithm is running on real numbers with infinite precision, which is not true when implemented on finite computers (our paper focuses on this gap).  

\subsection{Differential Privacy in Distributed Setup}\label{sec:dist_dp}

The standard definition of DP assumes a trusted server that collects all the data and runs the DP algorithm. In a distributed setup, the $m$ data owners can instead run an MPC protocol themselves and send the output to an analyst. Their number is large in practice, so a more practical design delegates the computation to $N$ untrusted servers. Following previous works~\cite{wei2023securely,fu2024benchmarking,roth2019honeycrisp,roth2020orchard}, we consider a distributed system that includes the following parties.
\begin{itemize}[leftmargin=*]
    \item {\bf $m$ users.} Each of the $m$ users secret-shares its data record (e.g., salary) to $N$ servers. All the data records submitted by the users form an underlying dataset $D$ protected by the system's DP guarantee.
    \item {\bf $N$ servers.} After receiving the secret shares from users, $N$ servers jointly run an MPC protocol $\Prot_f$, which contains the subroutines that compute the targeted query $f(D)$ (e.g., the total salary of some users), generate random noise, and add the noise to the query $f(D)$.
    \item {\bf An analyst.} The analyst can choose which query $f$ to compute on the specified group of users' records and the privacy budget $\epsilon$ to consume. The servers reconstruct the noisy query $f^*(D)$ and send it to a data analyst that conducts the subsequent data analysis tasks.
\end{itemize}

% \mypara{Computational Differential Privacy} The standard definition of DP holds against any adversary with unbounded computational power. However, in the distributed setup using the MPC protocol, the adversary's computational power is often assumed to be limited. To capture this, we can use the notion of computational differential privacy (CDP) proposed by Mironov et al.~\cite{mironov2009computational}, which only requires the DP guarantee to hold against polynomial-time adversaries. 

% \begin{definition}[($\epsilon$, $\delta$)-Computational Differential Privacy]
% \label{def:stand_comput_dp}
% Let $\epsilon \geq 0$ and $\delta \in [0, 1]$.
% A protocol $\Prot$ is ($\epsilon$, $\delta$)-computationally differentially private (($\epsilon$, $\delta$)-CDP) if for any neighboring datasets $D$ and $D'$, any non-uniform probabilistic polynomial-time adversary $\mathcal{A}$ and any non-uniform probabilistic polynomial-time distinguisher $\mathcal{D}$, there exists a negligible function $\text{negl}(\cdot)$, such that for all $\lambda$,
% \begin{align*}
%     & \Pr\left[\mathcal{D}\left({\sf View}^{\Prot}_\mathcal{A}(1^\lambda, D)\right) = 1 \right] \\
%       \leq & e^\epsilon \Pr\left[\mathcal{D}\left({\sf View}^{\Prot}_\mathcal{A}(1^\lambda, D')\right) = 1 \right] + \delta + \text{negl}(\lambda),
% \end{align*}
% where ${\sf View}_\mathcal{A}$ denotes the view of the adversary $\mathcal{A}$.
% \end{definition}

\section{Threat Model}\label{sec:threat_model}

We consider an honest-but-curious adversary who corrupts the data analyst to receive the output of protocol. The DP mechanism built in the protocol aims to protect the information about whether a record is present or absent in the input dataset. Hence, our threat model builds on this standard guarantee of DP. Specifically, we assume the adversary:

\begin{itemize}[leftmargin=*]
    \item observes the DP-protected output $f^*(D_b)$ of the protocol,
    \item knows two neighboring datasets $D_0$ and $D_1$ that differ in one record (consistent with the adding/deleting/replacing neighbor model of DP) and tries to guess which of the two datasets produced $f^*(D_b)$, i.e., the value of $b$. 
    \item knows how the protocol is implemented but does not know the randomness it uses.\footnote{For accountability, many deployed DP libraries are open-sourced~\cite{tfprivacy,googledp,opacus,diffprivlib}. Most of the victim systems in this paper are also open-sourced.}
\end{itemize}

By definition, the DP guarantee must hold even against an adversary who knows both $D_0$ and $D_1$: the two output distributions stay $(\epsilon,\delta)$-indistinguishable, which gives DP the ability to defend against membership inference~\cite{dwork2006differential}. Knowing the neighboring pair is therefore the definitional assumption rather than a strong one, and it is the threat model of many prior attacks on DP mechanisms~\cite{ilvento2020implementing,mironov2012significance,jin2022we,chourasia2026auditing,holohan2024securing}.

\section{Vulnerability in Sample-and-scale}~\label{sec:attack-uniform}
In this section, we highlight the vulnerabilities in existing continuous sampling protocols~\cite{gu2025dp,eigner2014differentially,pentyalacaps,pentyala2022training,roth2020orchard,roth2019honeycrisp}.

\subsection{Targeted Protocols Description}\label{sec:insecure-prot}

Although the protocols we attack can generate noise from different distributions (typically Laplace or Gaussian), they share a similar workflow as shown in Algorithm~\ref{alg:victim_protcol}.
Specifically, the protocols start by generating a secret share $\secret{u}$, where $u$ is a uniformly random number in $(0,1]$. Then, they use $\textsc{NonLinear}(\secret{u})$ to transform $\secret{u}$ to a secret-shared number from the standard Laplace / Gaussian distribution. For example, in~\cite{eigner2014differentially}, the standard Laplace sample is obtained by evaluating the well-known inverse CDF of Laplace distribution:  $\secret{r} \gets \secret{{\sf sign}} \times \ln (\secret{u})$,
where $\ln()$ is a protocol that computes the secret-shared natural logarithm of $u$ and ${\sf sign}$ is a uniformly random number drawn in $\{-1, 1\}$. The implementations of the protocol $\textsc{NonLinear}()$ and the supported types of query $f$ in existing works have been given in Table~\ref{tab:vulnerable}.

\begin{algorithm}[t]
\caption{Victim Protocol Workflow~\cite{gu2025dp,eigner2014differentially,pentyalacaps,pentyala2022training,roth2019honeycrisp,roth2020orchard}}
\label{alg:victim_protcol}
\begin{algorithmic}[1]
\Statex \hspace{-\algorithmicindent}\textbf{Input:} the public noise scale $s$, secret-shared query $\secret{f(D)}$
\Statex \hspace{-\algorithmicindent}\textbf{Output:} a secret share noise sample from Laplace / Gaussian distribution.
\State \textbf{// Sample a noise from the standard distribution}
\State $\secret{u} \gets {\sf Rand}_{(0,1]}()$ 
\State $\secret{r} \gets \textsc{NonLinear}(\secret{u})$  \Comment{$u \sim \lap(1)$ or $u \sim \gauss(1)$} \label{algl:non-linear}

\State \textbf{// Scale the noise by public parameter $s$}
\State \textcolor{red}{$\secret{\eta} \gets s \times \secret{r}$} \Comment{$\eta \sim \lap(s)$ or $\eta \sim \gauss(s^2)$} \label{algl:end_lap}
\State \textbf{// Perturb the query $\secret{f(D)}$ and reveal}
\State $\secret{f^*(D)} \gets \secret{f(D)} + \secret{\eta}$

\State \Return ${\sf Reveal}(\secret{f^*(D)})$ 
\end{algorithmic}
\end{algorithm}

\subsection{Fixed-point Attack Implementation}\label{sec:attack_impl}

\mypara{Warmup: special scaler format $s=2^t$} We now describe an intuitive case with the scaler $s$ in Algorithm~\ref{alg:victim_protcol} set to $s=2^t$, where $t\in \mathbb{N}^+$. Suppose the standard noise sample output from $\textsc{NonLinear}()$ (Line~\ref{algl:non-linear}) is $r=1.6875$ and the public scaler $s=4$. Under a fixed-point representation system with $n=4$ integer bits and $l=4$ fractional bits, multiplying $s$ and $r$ gives:
\[
\eta \;=\; s\cdot r \;=\; 6.75 \;=\; 0110.1100_2 .
\]

According to the arithmetic rule of fixed-point representation, the last two bits in the scaled noise are always ``$00$''.
Following the execution of Algorithm~\ref{alg:victim_protcol}, after receiving the noised query $f^*(D_b)$ from the protocol, the adversary can simply observe the last two bits and compare them with those in the queries $f(D_0)$ and $f(D_1)$ from the neighboring datasets. For example, assuming $f(D_0)=6.625$ and $f(D_1)=6.6875$, adding the ``flawed'' noise $\eta$ to them yields:
\[
\begin{aligned}
f^*(D_0)  &= 0110.10{\underline{10}}_2 + 0110.1100_2 
        = 1101.01{\underline{10}}_2,\\
f^*(D_1)  &= 0110.10{\underline{11}}_2 + 0110.1100_2 
        = 1101.01{\underline{11}}_2.
\end{aligned}
\]

Since the generated noise always contributes $00$ to the last two fractional positions, the last two bits (underlined) of the noisy query remain exactly the same as those of the private query result. To put it a bit more generally, for a scaler of format $s=2^t$, the last $t$ bits in the noised query $f^*(D)$ would remain the same as those in the original query $f(D)$.

The leakage itself, however, does not rely on the special format $s=2^t$. Under a general scaler the leakage no longer shows up as a direct zero padding, and instead it takes the form of a sparse output set. Next, we discuss the case of $s\geq 2$.

% Such a vulnerability is independent from the computation inside $\textsc{NonLinear}()$, i.e., even though $\textsc{NonLinear}()$ can provide a standard noise sample that exactly rounds into current fixed-point representation, as long as the scaling is performed in fixed point representation
\begin{figure}[t]
\centering
\includegraphics[width=\columnwidth]{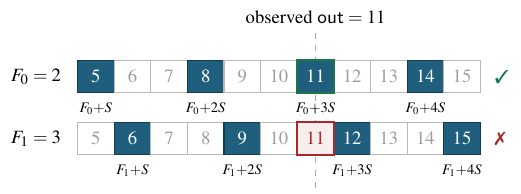}
\caption{Reachable outputs under the two neighboring datasets, shown in the integer encoding with an integer scaler $s=3$. A cell holds one representable value, and $F_b=2^{l}f(D_b)$ is the encoded query. The filled cells are the values $F_b+sR$ for $R\in\mathbb{Z}$, and the empty cells are values that the protocol cannot output. The two datasets reach disjoint sets here, so an output of $11$ rules out $D_1$.}
\label{fig:output}
\end{figure}

\mypara{Move to general case $s\geq 2$} Motivated by the above special case, we now consider a scaler $s$ of more general format. The warmup leans on the shift structure of $s=2^{t}$, which a general scale drops. Keeping the sample $r=1.6875$ from the warmup and changing the public scaler to $s=3$, under the same fixed-point system, multiplying $s$ and $r$ gives $\eta=s\cdot r=5.0625=0101.0001_2$,
whose last bit is $1$. For another example with $r=1.75$, the multiplication gives $\eta=5.25=0101.0100_2$, which ends with two zeros. The number of trailing zeros moves with $r$, so the adversary has no fixed position to read.

However, what survives here for every $s\geq 2$ is a statement about the values that the scaled noise $\eta=sr$ can reach. Write $R=2^{l}r$ for the equivalent integer encoding of the standard sample. Fixed-point multiplication computes the integer encoding of $\eta$ as $H=\lfloor sR\rfloor$, so $H$ always lies in a set of integers
\[
    \Lambda_s=\{\lfloor sR\rfloor\;:\;R\in\mathbb{Z}\},
\]
where two neighboring elements of $\Lambda_s$ differ by $\lfloor s\rfloor$ or $\lceil s\rceil$. Thus, for $s\geq 2$, the set $\Lambda_s$ leaves out at least half of the integers. For $s=3$ we get $\Lambda_3=\{0,3,6,9,\ldots\}$ and for $s=2.5$ we get $\Lambda_{2.5}=\{0,2,5,7,10,12,\ldots\}$. Leveraging the fact that $\Lambda_s$ is sparse and cannot cover all the possible values in the fixed-point representation system, adding $\eta$ to two different queries $f(D_0)$ and $f(D_1)$ leads to two output sets that the adversary can tell apart. Figure~\ref{fig:output} shows this separation in the integer encoding. Therefore, to infer whether the underlying dataset is $D_0$ or $D_1$, the adversary can subtract $f(D_0)$ and $f(D_1)$ from the output noisy query $f^*(D_b)$ respectively and see which results' integer encoding falls in the supported output set $\Lambda_s$. We formalize this procedure in Algorithm~\ref{alg:guess}.

\begin{algorithm}[t]
\caption{Guessing Algorithm}
\label{alg:guess}
\begin{algorithmic}[1]
\Statex \hspace{-\algorithmicindent}\textbf{Input:} the noisy query $f^*(D_b)$, the public noise scale $s$, and the two candidate queries $f(D_0)$, $f(D_1)$
\Statex \hspace{-\algorithmicindent}\textbf{Output:} a bit $b'\in\{0,1\}$ guessing the underlying dataset
\State $\mathsf{OUT} \gets 2^{l}f^*(D_b)$ \Comment{Move to the integer encoding}
\For{$i\in\{0,1\}$}
    \State $H_i \gets \mathsf{OUT}-2^{l}f(D_i)$ \Comment{Derive the noise $D_i$ needs}
    \State $c_i \gets \mathbbm{1}\big[\lceil H_i/s\rceil<(H_i+1)/s\big]$ \Comment{$c_i=1$ iff $H_i\in\Lambda_s$} \label{algl:membership}
\EndFor
\If{$c_0\neq c_1$} \Comment{Keep the reachable candidate}
    \State \Return $b'=i$ with $c_i=1$
\Else \Comment{Fallback to maximum likelihood}
    \State \Return the $b'=i$ that minimizes $|H_i|$
\EndIf
\end{algorithmic}
\end{algorithm}

Because fixed-point addition is exact, the released $f^*(D_b)=f(D_b)+\eta$ only shifts $\Lambda_s$ while keeping its step size. The true candidate therefore always passes the test in Line~\ref{algl:membership}. The test itself is efficient, since $H\in\Lambda_s$ holds exactly when the interval $[H/s,(H+1)/s)$ contains an integer. When both candidates pass, the membership test carries no information and the algorithm degenerates to the ordinary maximum likelihood rule, which returns the candidate closest to the release. The guess is therefore never worse than the one a correctly implemented mechanism already allows.

The success rate is decided by two quantities: the scale $s$ and the gap $d=2^{l}|f(D_0)-f(D_1)|$ between the two queries in the fixed-point encoding. Since the true candidate always passes the membership test, the algorithm degenerates to maximum likelihood only when the wrong candidate passes it as well, which requires $H\pm d$ to fall in $\Lambda_s$ in every released element when the query is a vector. From the results in Section~\ref{sec:attack-experiments}, under randomly chosen neighbouring dataset, such a degradation happens with very small probability, which is the reason why our attack achieves a near-$100\%$ success rate.

\mypara{Remark on the guessing algorithm} Our guessing algorithm only uses the fixed-point arithmetic rule, and it never cracks the cryptographic primitives that evaluate the sampling circuit nor the inner $\textsc{NonLinear}()$ implementation. The root cause is that multiplying the scaler $s$ with a standard sample $r$ from $\lap(1)$ or $\gauss(1)$ does not produce a sample from $\lap(s)$ or $\gauss(s^2)$ on $\mathbb{F}_{n,l}$. The leakage therefore persists even when $\textsc{NonLinear}()$ returns a sample whose distribution matches the exact rounding of the standard noise onto the grid.

% When the generated noise has $0$ in the last fractional bit, the last bit of $\mathsf{out}=f(D_b)+r^*$ is exactly that of $f(D_b)$, so $\mathsf{Guess}_{\mathsf{scale}}$ outputs $b$. In Figure~\ref{fig:result_noise}, we show the distribution of the last two bits for the concrete case $s=4$; the attack only needs one exposed bit. Our finding is that the last two bits are kept with about $60\%$ probability, while the expected probability should be $25\%$. As a result, with $95\%$ probability, $5$ samples will be sufficient for the adversary to determine whether the underlying dataset is $D_0$ or $D_1$.

\section{Experiments: Attacking Secure Aggregation}\label{sec:attack-experiments}

In this section, we run Algorithm~\ref{alg:guess} against two DP secure aggregation systems~\cite{roth2020orchard,gu2025dp} to demonstrate the impact of the uncovered vulnerability.

% \begin{table}[t]
% \centering
% \footnotesize
% \setlength{\tabcolsep}{4pt}
% \begin{tabular}{@{}>{\raggedright\arraybackslash}p{0.24\columnwidth}
%                    >{\raggedright\arraybackslash}p{0.38\columnwidth}
%                    >{\raggedright\arraybackslash}p{0.29\columnwidth}@{}}
% \toprule
% \textbf{System} & \textbf{Specified noise sampler} & \textbf{Representation} \\
% \midrule
% Honeycrisp (SOSP'19)
%   & Scale $r\sim\lap(1)$ by $s$ with $s=\Delta_f/\epsilon$
%   & SCALE-MAMBA \texttt{sfix}, $l=20$, $k=41$ \\[2pt]
% Orchard (OSDI'20)
%   & Scale $r\sim\lap(1)$ by $s$ with $s=\Delta_f/\epsilon$
%   & SCALE-MAMBA \texttt{sfix}, $l=20$, $k=41$ \\[2pt]
% DP-BREM$^+$ (USENIX Sec'25)
%   & Scale $r\sim\gauss(1)$ from Box--Muller by $s$
%   & Arithmetic circuit over field $\mathbb{F}_{2^k}$ \\
% \bottomrule
% \end{tabular}
% \caption{The systems we attack~\cite{roth2019honeycrisp, roth2020orchard,gu2025dp} for demonstration. Every paper specifies a sampler that scales a standard sample by a public constant. }
% \label{tab:victim-impl}
% \end{table}

\mypara{Implementation}
Orchard~\cite{roth2020orchard} supports 17 types of DP queries in the distributed setup and samples Laplace noise with the sample-and-scale protocol. DP-BREM$^+$~\cite{gu2025dp} is a DP federated learning system whose clients jointly draw a standard Gaussian sample and scale it by $R\sigma$ before adding it to the aggregated momentum of each training round. Inspecting the open-sourced implementation, we found that both systems leave out the sampler their paper specifies. Specifically, Orchard's compiler emits a \texttt{laplace\_fx} call that the prototype defines as adding a single random bit,\footnote{\scriptsize\url{https://github.com/edoroth/orchard/blob/e7cbed7/mpc_files/Programs/count_mean_sketch/count_mean_sketch.mpc\#L3-L4}} and DP-BREM$^{+}$ ships a plaintext PyTorch simulation that draws the noise with \texttt{torch.normal}.\footnote{\scriptsize\url{https://github.com/xiaolangu/DP-BREM/blob/ed4675c/train.py\#L236}} We therefore implement each sampler strictly following the description in the paper. For the fixed-point representation $\mathbb{F}_{n,l}$ in both systems, we set $n=21$ bits in the integer part and $l=20$ bits in the fractional part, following Orchard's default configuration (fixed-point format is left unspecified in DP-BREM$^{+}$~\cite{gu2025dp} and its code).

\mypara{Attack metrics} Following the threat model of Section~\ref{sec:threat_model}, the adversary knows two neighbouring datasets $D_0$ and $D_1$ and, from the values the system releases, guesses which one produced them. We report the success rate $\Pr[b'=b]$ over independent trials, where $b\in\{0,1\}$ is the secret index of protected dataset. Different from existing attacks, which allow the adversary to construct the neighbouring pair~\cite{mironov2012significance,jin2022we,ilvento2020implementing}, we obtain $D_1$ from a real dataset $D_0$ by removing one uniformly sampled record, so the gap $d$ between the two queries is whatever the data produces and the assumption on the adversary is weaker than that in previous works.

\mypara{Vectorized release} Both systems release a vector rather than a single value, such as the per-cluster sums of a $k$-means round in Orchard or the aggregated gradient of a training step in DP-BREM$^{+}$, so the adversary applies Algorithm~\ref{alg:guess} to each coordinate. The test is one-sided, since the noise the true dataset needs is reachable in every coordinate, so a candidate is discarded as soon as one coordinate rejects it and the guess falls back to maximum likelihood only when both candidates survive. Appendix~\ref{app:vector-guess} gives this vector form as Algorithm~\ref{alg:guess-vec}.

\mypara{Baseline} We compare against maximum likelihood on its own, which returns the candidate closest to the release and uses no property of the number representation. This is the distinguishing power a correctly implemented mechanism already grants the adversary, a more honest reference than a rate of $50\%$, and since the reachability test only ever discards a provably wrong candidate, our attack is at least as good as it on every trial. The closest candidate is measured in $\ell_1$ distance for Laplace noise and $\ell_2$ distance for Gaussian noise.

\subsection{DP $k$-Means with Laplace Noise}\label{sec:practical_attack}

\mypara{Victim system} Orchard~\cite{roth2020orchard} compiles a query into an MPC protocol run by the servers, and we attack its $k$-means query with every parameter taken from Orchard's own artifacts. In this system, users hold two-dimensional points, and the query runs $k=3$ clusters over $5$ rounds,\footnote{\scriptsize\url{https://github.com/hengchu/cps-fuzz/blob/master/src/Examples.hs\#L183-L237}} each round releasing the $x$-sum, $y$-sum and count of every cluster, that is $9$ noisy values per round and $45$ over the query. Each sum is clipped to $1$, so the sensitivity is $\Delta_f=1$ and the Laplace scaler is $s=1/\epsilon$. Here $\epsilon$ denotes the budget of a single release, which is the parameter Orchard's own artifacts set, and we vary it across both of their choices, i.e. $\epsilon=1$ in the compiled query and $\epsilon=0.1$ in the simulation script. Removing one user changes three of the nine values that a round releases, so basic composition over the five rounds gives a total budget of $15\epsilon$. We draw $10^4$ users uniformly from the geographic box the simulation uses and normalise the coordinates to the unit square that the clip bound and the initial centroids imply. Honeycrisp~\cite{roth2019honeycrisp}, the predecessor of this design, supports only one hard-coded query, so we do not attack it separately.

\begin{figure}[t]
\centering
\includegraphics[width=0.9\columnwidth]{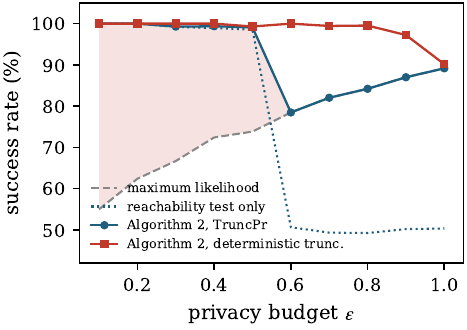}
\caption{Success rate against Orchard's $k$-means query, where $\epsilon$ is the per-release budget of Orchard's own artifacts and the total budget over the five rounds is $15\epsilon$. The shaded area is the advantage the attack gains from Algorithm~\ref{alg:guess} over the maximum likelihood baseline.}
\label{fig:kmeans-eps}
\end{figure}

\begin{figure}[t]
\centering
\includegraphics[width=0.9\columnwidth]{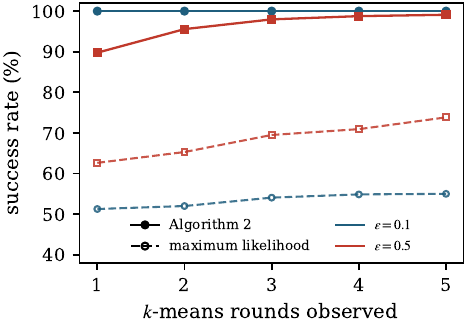}
\caption{Success rate as the adversary observes more rounds of the same $k$-means query, under \texttt{TruncPr}.}
\label{fig:kmeans-rounds}
\end{figure}

\begin{figure}[t]
\centering
\includegraphics[width=0.9\columnwidth]{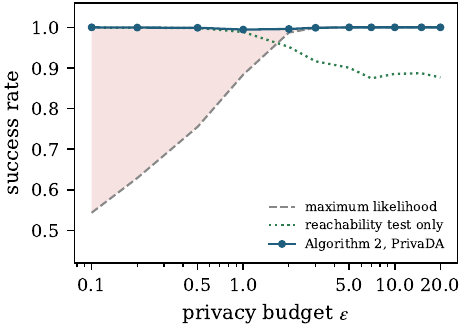}
\caption{Success rate against Orchard's $k$-means query when the noise is sampled by PrivaDA's floating-point mechanism \cite{eigner2014differentially} ($\beta=l=32$). The shaded area is the advantage of Algorithm~\ref{alg:guess} over the maximum likelihood baseline.}
\label{fig:kmeans-privada}
\end{figure}

\mypara{Results across $\epsilon$} Figure~\ref{fig:kmeans-eps} reports the success rate over $3\times 10^3$ queries per point, with the neighbouring dataset obtained by dropping one uniformly random user. The attack rules out a candidate dataset whose noise the scaled sampler could never have produced, so it needs the scaled noise to skip enough values. A larger scaler $s=1/\epsilon$ makes the noise skip more, so a smaller budget helps the attack, and the shaded area is what this adds on top of what differential privacy already concedes. Once the scaler drops below $2$ almost every value becomes reachable again, the test rules nothing out, and Algorithm~\ref{alg:guess} falls back to the baseline. The two truncation rules differ only in how fast this happens.The rule \texttt{TruncPr} used by SCALE-MAMBA~\cite{scalemamba} adds an additional rounding bit that fills in more values and loses its edge once $s<2$, i.e. $\epsilon>0.5$, whereas deterministic truncation keeps working down to $s=1$, i.e. up to $\epsilon=1$. The attack thus succeeds on essentially every query for $\epsilon\leq 0.5$ under either rule, which is the regime differential privacy is meant to protect.

\mypara{Rounds Accumulation} Since removing one user changes the three released values of the cluster that user falls in, $15$ out of the $45$ values carry a nonzero gap. Figure~\ref{fig:kmeans-rounds} shows what the adversary gains from keeping observing the output noisy queries. At a tight budget a single round already decides. Because in a single observation of 15 values, the Algorithm~\ref{alg:guess} may still fall back to maximum likelihood estimation. As the round increase, the adversary can observe more values to rule out the candidate query. Our attack approaches certainty by the fifth observation while the baseline is still far behind.

 \mypara{Floating-point variant} PrivaDA~\cite{eigner2014differentially} is the one protocol in Table~\ref{tab:vulnerable} that scales in floating point and converts to fixed point only at the end. We instantiate its sampler inside the same Orchard $k$-means attack, with every parameter fixed to its paper: a significand of $\beta=32$ bits, a $9$-bit exponent, and a fixed-point output with $l=32$ fractional bits. Floating point spaces its values by one unit in the last place, which grows with the magnitude, so the final conversion starts to skip grid points once $|s\cdot r|\geq 2^{\beta-l}$, i.e. once $|s\cdot r|\geq 1$ under PrivaDA's own $\beta=l=32$. This condition constrains the realized sample instead of the public scaler, which is where this variant departs from the fixed-point one. In fixed point a scaler $s\leq 1$ already rules the attack out, whereas here every sufficiently large draw leaks for any $s$. Figure~\ref{fig:kmeans-privada} varies the per-release budget from $\epsilon=0.1$ to $\epsilon=20$. The attack stays above $99\%$ over the whole range and reaches $100\%$ against a baseline of $54.3\%$ at $\epsilon=0.1$. Its gain closes beyond $\epsilon\approx 3$ only because the noise becomes small enough for the baseline itself to reach $100\%$. 

\subsection{DP Federated Learning}\label{sec:fl-attack}

\begin{figure}[t]
\centering
\includegraphics[width=0.9\columnwidth]{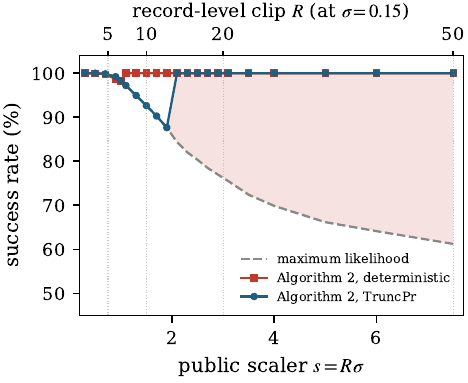}
\caption{Attacking DP-BREM$^+$ on MNIST as the scaler $s=R\sigma$ grows (larger per-round noise) at their clip configuration $R\in\{5, 10, 20, 50\}$. The shaded area is the advantage of Algorithm~\ref{alg:guess} over the maximum likelihood baseline.}
\label{fig:fl-sweep}
\end{figure}

\begin{figure}[t]
\centering
\includegraphics[width=0.9\columnwidth]{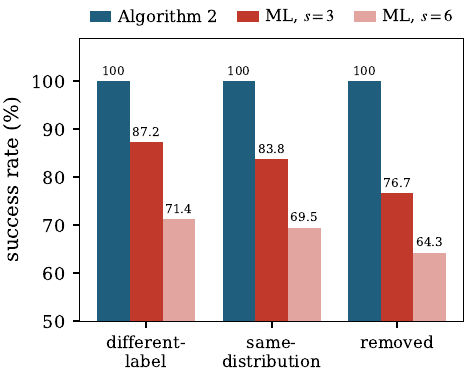}
\caption{Success rate as the neighbouring pair ranges from easy to hard. The attack is flat because the reachability test ignores the gap magnitude, whereas the baseline reads it.}
\label{fig:fl-canary}
\end{figure}

\mypara{Victim system} DP-BREM$^+$~\cite{gu2025dp} removes the trusted server by having the clients jointly generate the Gaussian noise that protects the aggregated momentum of a training round. We follow the released code and the reported hyper-parameters, i.e. the MnistCNN model with $26{,}010$ parameters, $100$ clients, record sampling rate $p=0.05$ and record-level clipping bound $R$. Each sampled record's gradient is clipped to $R$ and the clipped gradients are summed, so the sensitivity is $R$ and the noise scale is $s=R\sigma$. In each trial we take one client's batch, remove one uniformly sampled record from it, and simulate the adversary to test the two candidate aggregates of a single released round.

\mypara{Results across $s$} Different from the $k$-means query, one release from DP-BREM$^{+}$ is a $26{,}010$-dimensional vector of noisy momentum, and removing a record perturbs essentially all of its coordinates, so the adversary aggregates that many one-sided tests. Figure~\ref{fig:fl-sweep} varies $s=R\sigma$ at the paper's clip parameter. The paper specifies arithmetic circuits over $\mathbb{F}_{2^k}$ without fixing the truncation rule, so we implement both deterministic truncation and \texttt{TruncPr}. The reachability test switches on as a sharp step, at $s\geq 1$ under deterministic truncation and $s\geq 2$ under \texttt{TruncPr}, and above the step it rejects the wrong dataset on thousands of coordinates, which makes the attack certain. Maximum likelihood instead starts near certainty when the noise is tiny and decays to $61\%$ as $\sigma$ grows. The two are complementary, so the attack stays above $87\%$ everywhere and reaches $100\%$ for every $s\geq 2$. Crucially, the privacy budget is set by $\sigma$ alone and does not depend on $R$, so every configuration in the paper's own record-clip choices $R\in\{5,10,20,50\}$ carries the same total budget $\epsilon\approx1$, yet raising $R$ to reduce the clipping bias moves $s=R\sigma$ from $1.5$ to $3$ and $7.5$ and breaks the protocol under either truncation rule (Figure~\ref{fig:fl-sweep}). Although aggressively decreasing the clipping bound $R$ to narrow the noise scale seems to be a feasible mitigation to our attack, in practice, doing so will truncate the real gradient/momentum and lead to a loss of utility~\cite{abadi2016deep,andrew2021differentially}.

\mypara{Robustness to the neighbouring selection} Now we switch the format of candidate dataset pair in three ways, in decreasing order of how much they can challenge the released momentum of DP-BREM$^+$:
\begin{itemize}[leftmargin=*,itemsep=1pt,topsep=1pt]
    \item \textit{Different-label record:} one record is replaced by a record whose label is absent from the batch, so the two momentum vectors are in very different directions.
    \item \textit{Same-distribution record:} one record is replaced by another drawn from the same distribution, so vectors become closer.
    \item \textit{Removing a record:} one random record is dropped with no replacement, which produces the closest vector pairs.
\end{itemize}
The floating-point attack of Jin et al.~\cite{jin2022we} considers the first two setups, and its success rate falls from up to $92.6\%$ on the different-label record to $50\%$ on the same-distribution one (reducing to random guess), since a floating-point attack needs the two gradients to lie far apart. As shown in Figure~\ref{fig:fl-canary}, our attack instead stays at $100\%$ across all three cases, including the weakest removed-record pair, while the max likelihood (ML) baseline declines as the pair gets closer. The reason is that our attack fails only if the two candidate output momenta differ by an exact multiple of the grid step $s\,2^{-l}$, on every one of the $26{,}010$ coordinates, which is nearly impossible in practice. Fixed-point sampling is therefore strictly more vulnerable than floating-point, since the leakage needs only a nonzero gap between gradients rather than a large one.

\section{Securing Continuous Sampling Protocols}\label{sec:securing}

\subsection{Pyrrhic Fixes}\label{sec:pyrrhic}

Before giving our final secure sampling protocol, we start by describing some fixes which are direct but pyrrhic wins (feasible but inefficient).

\mypara{Aligning query $F$ and $\Lambda_s$}
The attack in Algorithm~\ref{alg:guess} succeeds for a single reason: the query lives on a finer domain $\mathbb{F}_{n,l}$ than the scaled noise can reach. In the integer encoding, the noise $H=\lfloor sR\rfloor$ lands on the grid $\Lambda_s$ of spacing $\approx s$, while the query $F=2^{l}f(D)$ ranges over the full integer domain, which is $s$ times finer. If the query and the noise instead shared the same grid, both noisy outputs would fall in the same reachable set. We first quantify this gap, which is the only thing an aligned mechanism must close.

\begin{lemma}\label{lem:bit-budget}
A standard sample of $l'$ fractional bits scaled by a public $s$ and rounded to the $b$-bit grid reaches every value of that grid if and only if $l'\ge b+\lceil\log_2 s\rceil$.
\end{lemma}

The proof is given in Appendix~\ref{app:bit-budget}. Scaling thus opens a $\lceil\log_2 s\rceil$-bit gap between noise and query. The two fix strategies below close this gap from two sides.

\mypara{Fix 1: rounding query $F$} This fix strategy brings the query down to the noise grid $\Lambda_s$. We write $\round_{\Lambda_s}(x)$ for the map that returns the element of $\Lambda_s$ closest to $x$. The fix rounds the query before adding the noise as follows:
\[
    \round_{\Lambda_s}(\mathsf{OUT})=\round_{\Lambda_s}(F)+H \qquad(H\in \Lambda_s).
\]
This is identical to rounding the \emph{released} noisy output $\mathsf{OUT}=F+H$ to $\Lambda_s$, which is similar to the idea of the snapping mechanism~\cite{mironov2012significance}. Either way every output lands in $\Lambda_s$ regardless of the dataset, so the reachability test of Algorithm~\ref{alg:guess} decides nothing. The price we pay in this fix is an output grid coarsened by $\lceil\log_2 s\rceil$ bits, which shows up as an additive error $O(s\,2^{-l})$ that grows with the scale, that is, a loss of utility.

\mypara{Fix 2: adding guard bits in noise} The second fix does the opposite and brings the noise grid up to the query grid. We sample the standard noise with $\lceil\log_2 s\rceil$ extra fractional bits. This shrinks the lattice step below one output unit (Lemma~\ref{lem:bit-budget} with $b=l$). After scaling, the noise now reaches every grid point and masks the query at full resolution. This fix keeps the utility. Instead, it pays in efficiency, because every operation runs on wider values. In an arithmetic circuit, one must either enlarge the ring of the whole protocol, or generate the noise in a larger ring and reduce it back. Both choices widen the sampler by a factor of $(l+\lceil\log_2 s\rceil)/l$.

\mypara{Dilemma} The two fixes are one construction seen from opposite sides. Unfortunately, neither of them can be proven secure for now, because both of them assume the standard sample $r$ output by $\textsc{NonLinear}()$ \emph{exactly} rounds the ideal continuous noise to $\mathbb{F}_{n,l}$, which touches another vulnerability independent of our attack: the standard noise output by an approximated $\textsc{NonLinear}()$ may already deviate from the claimed DP guarantee. To the best of our knowledge, no MPC protocol rounds these transcendental functions exactly, since $\ln(u)$ for Laplace and $\sqrt{-2\ln u_1},\cos(2\pi u_2)$ for Gaussian are only approximated with a bounded error~\cite{rathee2022secfloat,aliasgari2012secure,aly2019benchmarking}. Exact rounding is far more expensive, as \texttt{CR-LIBM}~\cite{daramy2003cr} evaluates polynomials on up to $118$ bits to produce a $32$-bit result, which would make $\textsc{NonLinear}()$, already the slowest part of the sampler (Section~\ref{sec:eval}), heavier still.

\subsection{Discrete Sampler for Continuous Noise}\label{sec:discrete-mechanism}
% Both of the above fixes either sacrifice the utility by rounding the query or increase the sampling cost by adding precision. More importantly, they require an inefficient non-linear function evaluation with exact rounding. 
In this section, we turn to the discrete mechanism directly on the bitwise granularity to sample noise under $\mathbb{F}_{n,l}$, which removes the scaling gap and the transcendental functions.

\mypara{From discrete to continuous} Note that every value in $\mathbb{F}_{n,l}$ is an integer multiple of the precision unit $2^{-l}$, so drawing a continuous noise on the grid is the same as drawing an integer and dividing it by $2^{l}$. Taking the Laplace noise as an example, let $Z\sim{\sf DLap}(t)$ be a discrete Laplace variable with $\Pr[Z=z]\propto e^{-|z|/t}$, and set the integer scale to $t=2^{l}s$. The fixed-point number $\eta=2^{-l}Z$ is supported on the grid, where for every integer $z$,
\[
    \Pr[\eta=z2^{-l}]\propto \exp\!\Big(-\frac{|z|}{2^{l}s}\Big)=\exp\!\Big(-\frac{|z2^{-l}|}{s}\Big),
\]
which is the density of $\lap(s)$ read on the grid. The integer $Z$ is therefore exactly the integer encoding of the noise, so the sampler produces $\eta$ directly on $\mathbb{F}_{n,l}$. A protocol only carries a finite number of bits, so we need a truncation parameter $\kappa$ that limits the range $Z \in [-2^{\kappa},2^{\kappa}]$, and we take the truncated distribution as the target that a secure sampler must produce. We define this target in Functionality~\ref{func:noise}.

\begin{functionality}[t]
\caption{$\Func[Lap](s, l, \kappa)$ for Laplace noise sampling}
\label{func:noise}
\smallskip
\noindent Samples a discrete Laplace variable $Z\sim{\sf DLap}_\kappa(2^{l}s)$, sets $\eta\gets 2^{-l}Z$, computes a fresh secret sharing $\secret{\eta}\in\mathbb{F}_{n,l}$, and sends each party its share.
\end{functionality}

Adding $\eta$ from $\Func[Lap]$ to $\secret{f(D)}$ yields a discrete Laplace mechanism that departs from the ideal $\lap(s)$ in two controlled ways. The grid rounds each value to a multiple of $2^{-l}$, which costs $O(2^{-l})$ in utility~\cite{canonne2020discrete}. The cap at $2^{\kappa}$ drops the tail of the distribution, which leaves a statistical distance of $O(e^{-2^{\kappa}/(2^{l}s)})$ from the untruncated distribution~\cite{fu2024benchmarking}. Both terms decay geometrically in $l$ and $\kappa$, so a modest budget makes them negligible. Inside $\Func[Lap]$ the sample $\eta$ never multiplies a public scaler and never evaluates a transcendental function, which closes the scaling gap of Section~\ref{sec:pyrrhic} and the exact-rounding dilemma together.

\mypara{Secured protocol construction} Now what remains is to efficiently draw $Z\sim{\sf DLap}_\kappa(2^{l}s)$ inside MPC. We follow the standard bitwise construction~\cite{dwork2006our,champion2019securely,wei2023securely}: a truncated geometric variable $G\sim{\sf Geo}(p)$ with $p=e^{-1/t}$ can be expressed as a concatenation of independently sampled biased coins, i.e., secret bits from Bernoulli distributions:
\[
    G_i\sim\bern(p[i]),\qquad p[i]=\frac{e^{-2^{i}/t}}{1+e^{-2^{i}/t}}.
\]
Therefore $G=\sum_{i} 2^{i}G_i$ comes from concatenating $\kappa$ independent coins. A discrete Laplace sample can be obtained by correcting $G$ into a two-sided noise~\cite{wei2023securely}. We abstract the coins as an ideal functionality $\Func[coin]$ that, given a public bias $p$, returns a secret share of $b\sim\bern(p)$. Algorithm~\ref{alg:dlap} describes the concrete protocol in the $\Func[coin]$-hybrid model. In our implementation, we instantiate the protocol~\cite{dwork2006our,wei2023securely} for $\Func[coin]$ with binary secret sharing in the semi-honest setting, where every party holds an XOR share of each bit and the AND gates are evaluated from Beaver triples. This instantiation tolerates a dishonest majority, which gives the following theorem.

\begin{algorithm}[t]
\caption{Discrete Laplace Sampler in the $\Func[coin]$-hybrid}
\label{alg:dlap}
\begin{algorithmic}[1]
\Statex \hspace{-\algorithmicindent}\textbf{Public parameters:} scale $s$, fractional bits $l$, truncation length $\kappa$, with $t=2^{l}s$
\Statex \hspace{-\algorithmicindent}\textbf{Output:} a secret-shared fixed-point noise $\secret{\eta}\in\mathbb{F}_{n,l}$ distributed as $\Func[Lap](s,l,\kappa)$
\State \textbf{// Truncated geometric from independent coins}
\For{$i=0$ to $\kappa-1$}
    \State $\secret{G_i}\gets\Func[coin](p[i])$ \Comment{$p[i]=e^{-2^{i}/t}/(1+e^{-2^{i}/t})$}
\EndFor
\State $\secret{G}\gets\sum_{i=0}^{\kappa-1}2^{i}\secret{G_i}$
\State \textbf{// Geometric $\to$ discrete Laplace}
\State $\secret{c}\gets\Func[coin](p_0)$; \ \ $\secret{b}\gets\Func[coin](1/2)$ \Comment{zero correction and sign}
\State $\secret{Z}\gets (1-\secret{c})\cdot(2\secret{b}-1)\cdot(\secret{G}+1)$
\State \Return $\secret{\eta}\gets 2^{-l}\secret{Z}$
\end{algorithmic}
\end{algorithm}

\begin{theorem}\label{thm:dlap-secure}
In the $\Func[coin]$-hybrid model, Algorithm~\ref{alg:dlap} securely realizes $\Func[Lap]$ against a semi-honest adversary corrupting up to $N-1$ parties.
\end{theorem}

The protocol for sampling Gaussian noise is analogous. It replaces the geometric variable with a geometric proposal and a rejection test~\cite{canonne2020discrete,wei2023securely}, and we defer its functionality and protocol to Appendix~\ref{app:dgauss}.

\subsection{Implementation and Improvement}\label{sec:impl}

The construction in Section~\ref{sec:discrete-mechanism} is correct and secure, but it calls $\Func[coin]$ once for every bit of the noise. An integer sampler of scale $s$ flips $O(\log s)$ coins~\cite{fu2024benchmarking,dwork2006differential,wei2023securely} to place the noise on the grid of $\mathbb{F}_{n,l}$, so a single sample already needs $O(l+\log s)$ coins. This cost strains existing implementations~\cite{fu2024benchmarking}. A constant-round backend such as Yao's garbled circuit or BMR incurs large total communication, while a secret-sharing backend needs a communication round for every secret-sharing multiplication, because MP-SPDZ~\cite{keller2020mp} does not batch the coin multiplications on its own. In fact, all of these coins are independent and admit parallel evaluation, which our implementation below exploits.

\mypara{Parallelization of coins flipping} In our implementation, we view the implementation of each $\Func[coin]^i$ as a circuit of $\mathcal{T}$ sequential AND gates evaluated on binary secret shares:
\begin{align*}
    \text{Coin}_i \gets \{{\sf AND}_{1,i}, \dots, {\sf AND}_{\mathcal{T},i}\}
\end{align*}
Since all the coins $\text{Coin}_1, \dots, \text{Coin}_{\kappa+l}$ are flipped independently, we can evaluate the $\mathcal{T}$-depth circuit of all the coins in parallel, which requires $O(\mathcal{T})$ online communication rounds.
Specifically, a noise sampler that produces $B$ noise samples needs to flip $B(\kappa+l)$ independent coins. Thus, for $j\in \{1, \dots, \mathcal{T}\}$, we orchestrate $B(\kappa+l)$ independent AND gates at the same depth $j$ into a single vectorized AND gate that communicates in constant rounds: 
 \begin{align*}
    \text{Round}_j \gets \{{\sf AND}_{j,1}, \dots, {\sf AND}_{j,{B(\kappa+l)}}\}
\end{align*}
In the end, flipping $B(\kappa+l)$ independent coins can be done in $O(\mathcal{T}B(\kappa+l))$ communication but only $O(\mathcal{T})$ online rounds. Such an optimization is crucial for multi-party computation in real-world WAN setting, where the latency of each round is high and dominates the total running time.

\mypara{From Boolean to arithmetic shares} The coin flipping circuit produces each bit as a Boolean share, whereas the query $\secret{f(D)}$ is arithmetic. We therefore append a single batched share conversion~\cite{escudero2020improved} that lifts all $B$ samples to arithmetic shares together, after which the noise is added to the query with local operations. This conversion needs a constant number of rounds for the whole batch of $B$ noise samples.

\mypara{Backend} We build the sampler directly on the low-level Boolean API of MP-SPDZ~\cite{keller2020mp}. Its high-level compiler exposes a parallel-loop interface, \texttt{for\_range\_parallel},\footnote{\url{https://mp-spdz.readthedocs.io/en/latest/Compiler.html}} that vectorizes independent iterations, but a batch of our size emits a large amount of bytecode instructions that compiling and loading them dominates the native batched operation. We therefore bypass the compiler and hand the backend the batched Boolean layers ourselves, which keeps us in control of the online schedule. The same schedule runs unchanged under a semi-honest or a maliciously secure backend, and a stronger backend changes only the preprocessing cost rather than the online round count.

\section{Experiments: Secured Sampler}\label{sec:secure-experiments}

% In this section, we evaluate the performance of our secured
% sampler in terms of efficiency, correctness, and utility.

\begin{figure*}[t]
\centering
\setlength{\abovecaptionskip}{3pt}
\includegraphics[width=\textwidth]{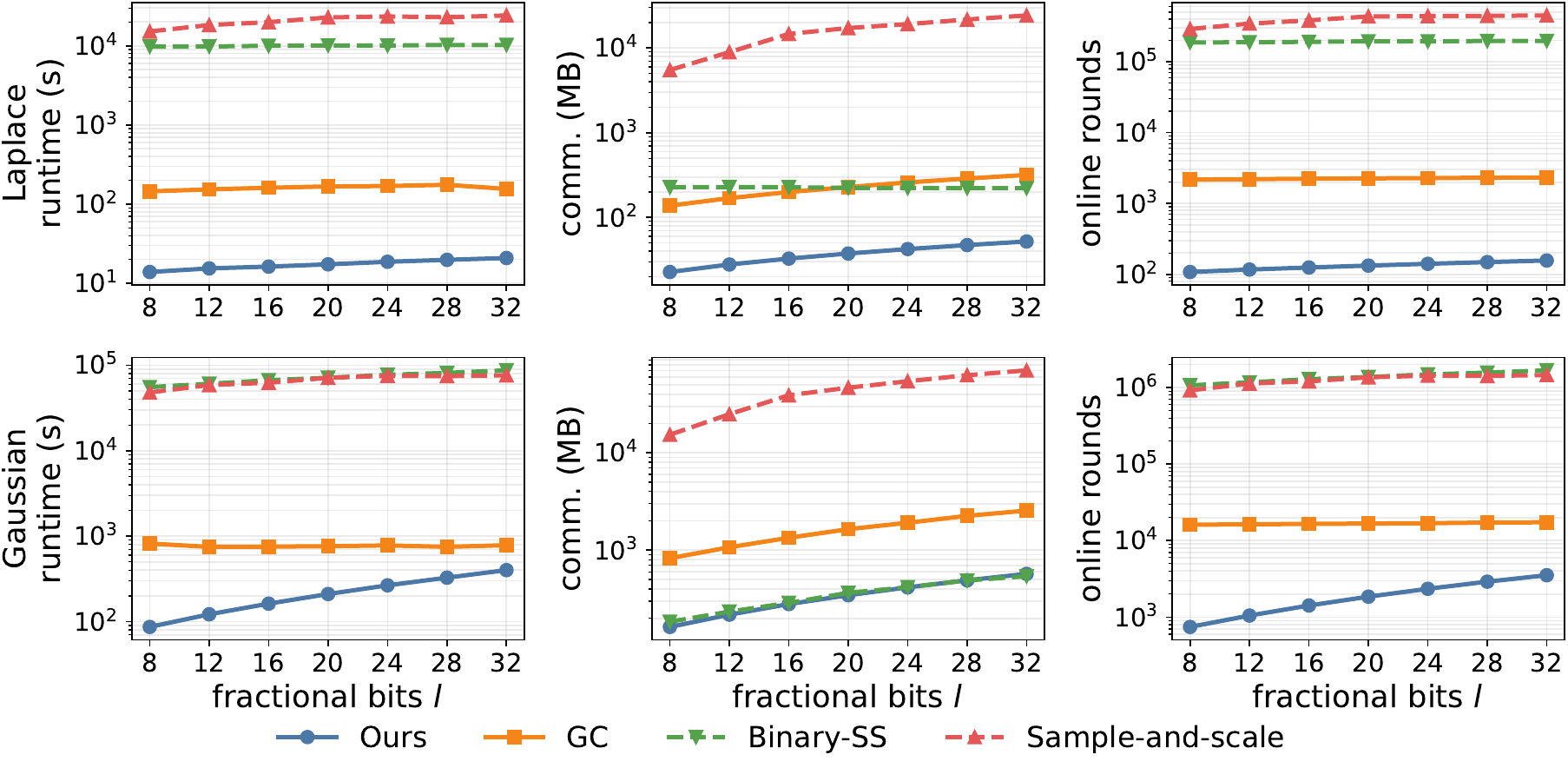}
\caption{Efficiency comparison over different fractional-bit configurations $l$, for $B=1{,}024$ samples, $\lambda=64$, scale $10\cdot 2^{l}$. GC and Binary SS are the same implementation of~\cite{fu2024benchmarking} executed with different backends \texttt{semi-bin-party.x} for secret sharing and \texttt{yao-party.x} for garbled circuits.}
\label{fig:eff-sweep}
\end{figure*}

\subsection{Efficiency}\label{sec:eval}

\mypara{Setup and evaluated protocols} All efficiency benchmarks run on a server with a 16-core AMD Threadripper PRO CPU and 256 GB RAM, under a WAN emulation with up to 1 Gbps bandwidth and up to 100 ms latency.
We consider the following protocols on the fixed-point grid for comparison, with the same security parameter configuration $\lambda=64$:
\begin{itemize}[leftmargin=*]
    \item {\it Insecure sample-and-scale in Section~\ref{sec:insecure-prot}.} We implement the samplers for the Laplace and Gaussian mechanisms using the MP-SPDZ~\cite{keller2020mp} framework. We use the framework's default approximated math library \texttt{mpc\_math} to implement $\textsc{NonLinear}()$ and generate standard noise. The scaling by $s$ uses the default multiplication of MP-SPDZ.

    \item {\it Existing discrete sampler implementation~\cite{fu2024benchmarking}.} We convert this implementation for discrete Laplace / Gaussian to support fixed-point $\mathbb{F}_{n,l}$ grid using Algorithm~\ref{alg:dlap} and \ref{alg:dgauss}. For every setting with scaler $s$, we instead use a scaler $s2^l$ to provide extra precision in the fractional part with $l$ bits.
    \item {\it Secure lookup table sampler~\cite{franzese2025secure}.} Instead of using coin flipping, this protocol rely on securely looking up a pre-computed table that contains all the values for a truncated discrete Gaussian distribution. Similarly, by setting scaler to $s2^l$, it can support the fixed-point $\mathbb{F}_{n,l}$ grid in the semi-honest, single corruption setup.
    \item {\it Our discrete sampler implementation.} We implement the discrete Laplace / Gaussian sampler in Algorithm~\ref{alg:dlap} and \ref{alg:dgauss} with the optimizations in Section~\ref{sec:impl}. 
\end{itemize}

\mypara{Comparison with baselines}
This paper focuses on the fixed-point grid $\mathbb{F}_{n,l}$, which introduces an additional parameter $l$ for the fractional part. In this evaluation, we fix the scaler $s=10$, vary the commonly used precision $l$ from $8$ to $32$, and compare the protocols in wall-clock runtime, communication and online rounds. We run all the protocols in the semi-honest two-party setting. For sample-and-scale, we run it with MP-SPDZ's \texttt{semi2k-party.x} backend over $\mathbb{Z}_{2^k}$. As for the implementation of~\cite{fu2024benchmarking}, since it is a binary protocol, we run it with \texttt{semi-bin-party.x} for secret sharing and \texttt{yao-party.x} for garbled circuits. Our sampler is also run with a two-party binary secret-sharing backend.

Figure~\ref{fig:eff-sweep} reports the costs as the precision $l$ grows. Our sampler lowers the runtime over the insecure sample-and-scale by up to $1{,}300\times$ for Laplace and $550\times$ for Gaussian, and over the secure baselines it is $2\times \sim 10\times$ faster than the garbled circuit and $217\times\sim 706\times$ faster than binary secret sharing~\cite{fu2024benchmarking}. Our communication stays close to binary secret sharing, yet we cut its online rounds by two to three orders of magnitude, which yields a smaller runtime in the WAN setup. The garbled circuit keeps its rounds almost flat in $l$ while its total communication is four to six times larger. Our sampler keeps a good balance between the number of rounds and the communication. We further compare against the secure lookup table sampler in~\cite{franzese2025secure} under the same setting of the Table~1 in their paper. We use the numbers reported in their paper. Our sampler is $1.5\times \sim 53\times$ faster on a LAN and matches them on a WAN, while it sends less data at two and four parties and more at sixteen. We show the concrete numbers and discuss their protocol in Appendix~\ref{app:lut}.

\subsection{Correctness and Utility}\label{sec:correctness}
Section~\ref{sec:discrete-mechanism} show that our discrete sampler securely realizes its ideal functionality. We now empirically complement this proof with an empirical study.

\begin{figure}[t]
\centering
\setlength{\abovecaptionskip}{3pt}
\includegraphics[width=\columnwidth]{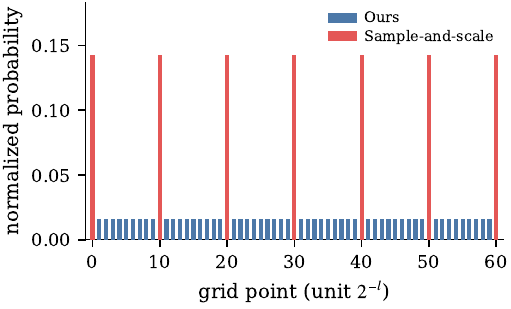}
\caption{Reachable values on a small window $[0,60]\cdot 2^{-l}$ of the fixed-point grid at $l=16$ and $s=10$ for the Laplace noise. Each bar is the probability of one grid point, normalized to unit mass inside the window.}
\label{fig:noise-coverage}
\end{figure}

\begin{table}[t]
\centering
\footnotesize
\begin{tabular*}{0.9\columnwidth}{@{\extracolsep{\fill}} c cc cc @{}}
\toprule
\multirow{2}{*}{\textbf{Precision}} & \multicolumn{2}{c}{\textbf{Laplace}} & \multicolumn{2}{c}{\textbf{Gaussian}} \\
\cmidrule(lr){2-3}\cmidrule(lr){4-5}
 & $D_n$ & KS $p$ & $D_n$ & KS $p$ \\
\midrule
$l=8$  & 0.014 & 0.50 & 0.014 & 0.50 \\
$l=16$ & 0.012 & 0.62 & 0.014 & 0.51 \\
$l=32$ & 0.012 & 0.58 & 0.013 & 0.55 \\
\bottomrule
\end{tabular*}
\caption{Kolmogorov--Smirnov test of our sampler against the discrete distribution of the ideal functionality, at scaler $s=10$ and precisions $l$. $D_n$ is the KS statistic, and the $p$-value is estimated by a parametric bootstrap over $2{,}000$ reference datasets of size $4{,}096$ and averaged over $20$ runs.}
\label{tab:ks}
\end{table}

\mypara{Noise distribution}
We first examine the distribution that our sampler produces. We fix the scaler $s=10$, draw samples at the precisions $l\in\{8,16,32\}$, and compare them against the discrete Laplace and Gaussian distributions of Functionality~\ref{func:noise} and \ref{func:gauss} with a one-sample Kolmogorov--Smirnov test, whose $p$-value comes from a parametric bootstrap that suits the discrete target. As shown in Table~\ref{tab:ks}, the test never rejects at the $0.05$ level, and the $p$-value stays near the $0.5$ that a correct sampler produces, so the samples stay consistent with the intended mechanism. Figure~\ref{fig:noise-coverage} then observes the probability that our protocol assigns to each grid point, in the window $[0,60]\cdot 2^{-l}$ of the grid at $l=16$, where adjacent grid points are spaced $2^{-l}$ apart and the probabilities are normalized to unit mass inside the window. Our sampler places mass on every grid point, while sample-and-scale reaches only the multiples of $s$ and leaves the points between them empty. The Laplace and Gaussian cases have similar distributions at this granularity, so we show only the Laplace one. These two results confirm empirically that our sampler is correct and secure against the attack we launched in Section~\ref{sec:attack_impl}.

\begin{figure}[t]
\centering
\setlength{\abovecaptionskip}{3pt}
\includegraphics[width=\columnwidth]{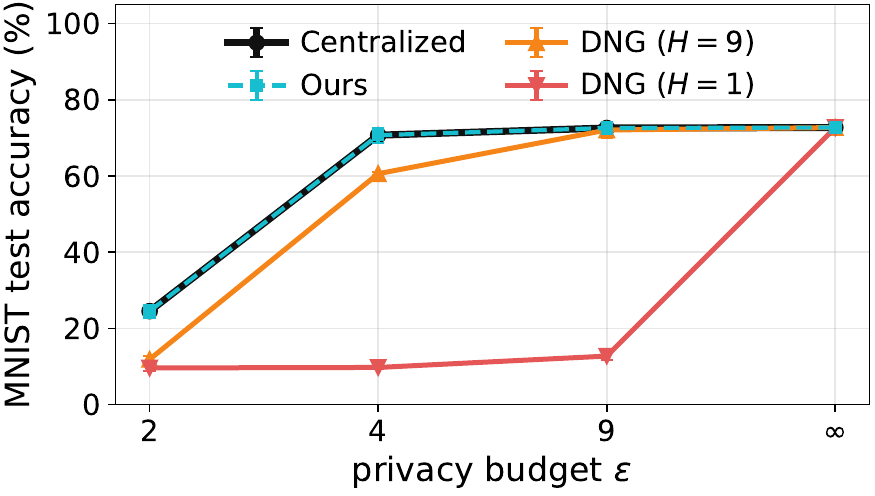}
\caption{MNIST test accuracy of the federated learning task of Gu et al.~\cite{gu2025dp} under different noise sources, over the privacy budget $\epsilon$ at $N=16$ parties. Our sampler overlaps the centralized mechanism, while distributed noise generation with $H$ honest parties loses utility as $H$ decreases.}
\label{fig:casestudy}
\end{figure}

\mypara{Case study: DP federated learning}
We follow the federated learning setup of Gu et al.~\cite{gu2025dp} on MNIST with $N=16$ parties and replace its Gaussian noise with our correct and secure sampler. We compare the utility with the distributed noise generation (DNG) baseline, which requires each party to locally sample Gaussian noise and then aggregate the shares in MPC. Since an adversary that controls $N-H$ parties can subtract their partial noise and weaken the guarantee~\cite{fu2024benchmarking}, a direct fix makes the $H$ honest parties' partial noise already sum to the target Gaussian, so each party enlarges its noise and the aggregate that the model receives grows to $N/H$ times the target variance. Figure~\ref{fig:casestudy} varies the number of honest parties $H$ and reports the MNIST test accuracy over the privacy budget $\epsilon$. Our sampler matches the centralized mechanism at every budget, while the distributed baseline loses accuracy as $H$ shrinks, and the dishonest-majority case $H=1$ stays unusable until the noise vanishes at $\epsilon=\infty$.

\section{Related Work}\label{sec:related}

\mypara{{DP in Finite Representations}}
Mironov~\cite{mironov2012significance} demonstrates that floating-point implementations of the Laplace mechanism in the standard library leak information through the least significant bits. Canonne et al.~\cite{canonne2020discrete} propose discretized mechanisms for DP (discrete Laplace/Gaussian) to address floating-point vulnerabilities. Jin et al.~\cite{jin2022we} further reveal timing side-channel vulnerabilities on discrete Gaussian as well as the floating-point implementation of the Gaussian mechanism. Holohan et al.~\cite{holohan2024securing} analyze the vulnerabilities in the normalization and rounding steps of floating-point DP mechanisms. Ilvento~\cite{ilvento2020implementing} shows that the floating-point implementation of the exponential mechanism can also be broken by a similar attack, and proposes base-2 DP to align the privacy budget with binary representations. More recently, Chourasia et al.~\cite{chourasia2026auditing} audit Apple's DP framework and identify multiple vulnerabilities caused by finite representation designs.

\mypara{Secure Sampling Protocols in MPC} Dwork et al.~\cite{dwork2006our} introduce discrete noise generation via secret sharing. Champion et al.~\cite{champion2019securely} formalize secure coin flipping for biased Bernoulli sampling, enabling bit-by-bit construction of discrete distributions. Wei et al.~\cite{wei2023securely} extend discrete Laplace sampling to discrete Gaussian sampling. Fu and Wang~\cite{fu2024benchmarking} benchmark these bitwise protocols in MP-SPDZ~\cite{keller2020mp}. Recently, Franzese et al.~\cite{franzese2025secure} replace the bitwise circuit with a secure lookup table, which supports any discrete distribution and scales well with the number of parties. 
Another line of work focuses on continuous noise generation~\cite{gu2025dp, eigner2014differentially, pentyalacaps, pentyala2022training, goryczka2015comprehensive, bohler2021secure,ruan2023private}, which ignores the vulnerabilities in fixed-point implementations that we analyze systematically. Keller et al.~\cite{keller2024secure} propose secure sampling for floating-point MPC, which still incurs high overhead from complex floating-point operations.

% \section{Discussions and Future Work}\label{sec:discuss}
% Our systematic analysis of continuous noise generation protocols suggests several directions for future research.
% \begin{itemize}[leftmargin=*]
%     \item {\it Floating-point sampling protocols.}  While most existing works focus on fixed-point representations prevalent in MPC frameworks, some applications (e.g., DP-SGD for large models) benefit from the dynamic range of floating-point arithmetic. However, all the normalization and rounding vulnerabilities in standard floating-point libraries are inherited when extending to MPC. Designing a secure floating-point mechanism in a DP framework is a promising future direction.
    
%     \item {\it Secure non-linear functions with exact rounding.}  Current polynomial approximations~\cite{rathee2022secfloat,aliasgari2012secure} in MPC cannot offer exact rounding.  
%     Achieving exact rounding for $\ln(u)$, $\sqrt{-2\ln u}$ and $\cos(u)$ in MPC would enable transformation sampling to satisfy DP over fixed-point outputs.  

%     \item {\it Better trade-offs between efficiency and precision.} Our analysis and evaluation show that sampling fixed-point noise with high precision would incur additional overhead. Bitwise sampling can achieve linear overhead scaling in fractional precision $l$ while the table lookup approach scales exponentially. In the future, it is interesting to explore new sampling protocols that can achieve better trade-offs between efficiency and precision.
% \end{itemize}

\section{Conclusion}\label{sec:conclude}

In this paper, we revisit the sample-and-scale paradigm that existing systems use to generate continuous noise in MPC. We show that scaling on the fixed-point grid confines the released query to a sparse and publicly known set. Since direct repairs remain costly, we instead sample the noise on the grid with parallel biased coins and prove the protocol secure. The resulting sampler matches the utility of the ideal continuous mechanism while achieving improvements in efficiency.

\bibliographystyle{plain}
\bibliography{refs}

\appendix

\section{Proof of Lemma~\ref{lem:bit-budget}}\label{app:bit-budget}
\begin{proof}
Encode the standard sample with $l'$ fractional bits, so $R=2^{l'}r\in\mathbb{Z}$; scaling by $s$ and rounding to the $b$-bit grid gives $H=\big\lfloor sR/2^{\,l'-b}\big\rceil$. As $R$ runs over consecutive integers, $sR/2^{\,l'-b}$ advances in steps of $s\,2^{-(l'-b)}$, so $\{H\}$ hits every point of the $b$-bit grid if and only if this step is at most one grid unit,
\[
    s\,2^{-(l'-b)}\le 1 \iff l'-b\ge \log_2 s \iff l'\ge b+\lceil\log_2 s\rceil .
\]
When $l'<b+\lceil\log_2 s\rceil$ the step exceeds one unit, so $\{H\}$ lands on a sublattice of density $2^{\,l'-b}/s<1$ whose coset within the grid is shifted by $f(D)$, which is the leak exploited in Section~\ref{sec:attack-uniform}.
\end{proof}

\section{Vectorized Guessing Algorithm}\label{app:vector-guess}

Algorithm~\ref{alg:guess-vec} is the vector form of Algorithm~\ref{alg:guess} used in Section~\ref{sec:attack-experiments}. It runs the reachability test on every coordinate of the released vector, keeps a candidate only when all of its coordinates are reachable, and degenerates to maximum likelihood when both candidates survive. The distance $\lVert\cdot\rVert$ in Line~\ref{algl:vec-ml} is $\ell_1$ for Laplace noise and $\ell_2$ for Gaussian noise.

\begin{algorithm}[h]
\caption{Vectorized Guessing Algorithm}
\label{alg:guess-vec}
\begin{algorithmic}[1]
\Statex \hspace{-\algorithmicindent}\textbf{Input:} the noisy query $f^*(D_b)\in\mathbb{R}^{q}$, the public noise scale $s$, and the two candidate queries $f(D_0)$, $f(D_1)\in\mathbb{R}^{q}$
\Statex \hspace{-\algorithmicindent}\textbf{Output:} a bit $b'\in\{0,1\}$ guessing the underlying dataset
\State $\mathsf{OUT} \gets 2^{l}f^*(D_b)$ \Comment{coordinate-wise integer encoding}
\For{$i\in\{0,1\}$}
    \State $c_i \gets 1$
    \For{$j\in\{1,\dots,q\}$}
        \State $H_{i,j} \gets \mathsf{OUT}_j-2^{l}f(D_i)_j$
        \State $c_i \gets c_i \wedge \mathbbm{1}\big[H_{i,j}\in\Lambda_s\big]$ \Comment{reject $i$ if any coordinate is unreachable}
    \EndFor
\EndFor
\If{$c_0\neq c_1$}
    \State \Return $b'=i$ with $c_i=1$
\Else
    \State \textbf{// Degenerate to maximum likelihood}
    \State \Return the $b'=i$ that minimises $\lVert \mathsf{OUT}-2^{l}f(D_i)\rVert$ \label{algl:vec-ml}
\EndIf
\end{algorithmic}
\end{algorithm}

\section{Security of the Discrete Sampler}\label{app:dlap-proof}
\begin{proof}[Proof of Theorem~\ref{thm:dlap-secure}]
Let $\mathcal{C}\subsetneq\{P_1,\dots,P_N\}$ be the corrupted set. We build a simulator $\mathcal{S}$ that, given only the corrupted parties' shares of the output of $\Func[Lap]$, reproduces their view of Algorithm~\ref{alg:dlap}. Every step of the algorithm is one of three kinds: (i) a call to $\Func[coin]$, whose only message to $\mathcal{C}$ is a secret share of a fresh bit and is simulated by handing $\mathcal{C}$ uniformly random shares; (ii) a local linear operation ($\sum_i 2^i\secret{G_i}$, the public multiplications, the final $2^{-l}$ scaling), which sends no message; and (iii) the two Boolean multiplications forming $(1-\secret{c})(2\secret{b}-1)(\secret{G}+1)$, each realized by consuming a Beaver triple, whose opened values are uniform given the corrupted shares and are simulated as such. Since no step reveals any secret beyond the shares $\mathcal{C}$ already holds, $\mathcal{S}$'s output is distributed identically to the real view, and the reconstructed sample equals the one $\Func[Lap]$ outputs by correctness of the coin-to-Laplace map. The claim follows.
\end{proof}

\section{Comparison with the Secure Lookup Table}\label{app:lut}

Franzese et al.~\cite{franzese2025secure} sample discrete noise from a secure lookup table instead of a bitwise circuit. The parties hold $\lambda$ tables whose length covers the support of the target distribution, and each lookup is evaluated on an encrypted one-hot vector of that length, which parallelizes well across parties. Their construction is generic over discrete distributions and works in the same semi-honest, all-but-one corruption setting as ours.
Table~\ref{tab:lut-comparison} compares the two samplers in the setting of their Table~1, which fixes a discrete Gaussian with $\sigma=967$, $B=32{,}768$ samples, and $\lambda=64$, under a WAN of $100$~ms latency and $1$~Gbps bandwidth. We take their numbers directly from their paper and measure ours under the same setting.

Our sampler is strictly faster on a LAN and matches theirs on a WAN. The reason is that their lookup runs under homomorphic encryption and is therefore computation bound, so its LAN and WAN times stay close. Ours is bound by the number of communication rounds instead, and the Gaussian sampler still needs $689$ online rounds after the optimizations of Section~\ref{sec:impl}, which the WAN latency then dominates.

\begin{table}[h]
\centering
\footnotesize
\begin{tabular*}{0.9\columnwidth}{@{\extracolsep{\fill}} l l cccc @{}}
\toprule
\multirow{2}{*}{\textbf{Metric}} & \multirow{2}{*}{\textbf{Method}} & \multicolumn{4}{c}{\textbf{Number of parties}} \\
\cmidrule(lr){3-6}
 & & \textbf{2} & \textbf{4} & \textbf{8} & \textbf{16} \\
\midrule
\multirow{2}{*}{LAN (s)}     & Ours & 0.0017 & 0.0017 & 0.0093 & 0.04 \\
                             & LUT  & 0.09 & 0.07 & 0.06 & 0.06 \\
\cmidrule(lr){1-6}
\multirow{2}{*}{WAN (s)}     & Ours & 0.10 & 0.11 & 0.11 & 0.24 \\
                             & LUT  & 0.10 & 0.08 & 0.08 & 0.25 \\
\cmidrule(lr){1-6}
\multirow{2}{*}{Comm.\ (MB)} & Ours & 0.13 & 0.76 & 3.54 & 15.10 \\
                             & LUT  & 0.72 & 1.91 & 3.30 & 11.59 \\
\bottomrule
\end{tabular*}
\caption{Comparison with the secure lookup table sampler of Franzese et al.~\cite{franzese2025secure}. LAN and WAN report the amortized time and the global communication per sample. The LUT rows are the numbers reported in~\cite{franzese2025secure}.}
\label{tab:lut-comparison}
\end{table}

This comparison fixes one distribution, and we do not include the lookup table in the precision comparison of Figure~\ref{fig:eff-sweep}. The reason is that its table length has to cover the support of the target distribution. Reading the noise on the fixed-point grid $\mathbb{F}_{n,l}$ multiplies that support by $2^{l}$, so the same $\sigma$ that needs $2^{16}$ entries in their evaluation needs roughly $2^{24}$ entries at $l=16$ and $2^{40}$ at $l=32$, which lies far outside the range their evaluation covers.

\section{Discrete Gaussian Sampler}\label{app:dgauss}
The Gaussian case mirrors the Laplace one. We target the analogous functionality $\Func[Gauss]$, which outputs a truncated discrete Gaussian read on the grid.

\begin{functionality}[t]
\caption{$\Func[Gauss](\sigma, l, \kappa)$ for Gaussian noise sampling}
\label{func:gauss}
\smallskip
\noindent Samples a discrete Gaussian variable $Z\sim{\sf DGauss}_\kappa(2^{l}\sigma)$, sets $\eta\gets 2^{-l}Z$, computes a fresh secret sharing $\secret{\eta}\in\mathbb{F}_{n,l}$, and sends each party its share.
\end{functionality}

Algorithm~\ref{alg:dgauss} gives the protocol. It follows the same template as Algorithm~\ref{alg:dlap}, which it reuses to draw a discrete Laplace proposal $Y$ at the integer scale $t$, and then accepts $Y$ with the probability the rejection sampler of~\cite{canonne2020discrete,wei2023securely} prescribes. The acceptance test needs no new primitive, because a secret integer $X$ decomposes into bits and $e^{-X/D}=\prod_i\big(e^{-2^{i}/D}\big)^{X_i}$, so one $\Func[coin]$ call per bit of $X$ suffices. A rejection loop would leak the trial count through its running time, hence the protocol draws a public number $M$ of proposals in parallel and obliviously selects the first accepted one. We set $M$ so that at least one proposal is accepted except with probability $\delta$, and $\delta$ enters the truncation budget of $\Func[Gauss]$ in the same way as $\kappa$.

\begin{algorithm}[t]
\caption{Discrete Gaussian Sampler in the $\Func[coin]$-hybrid}
\label{alg:dgauss}
\begin{algorithmic}[1]
\Statex \hspace{-\algorithmicindent}\textbf{Public parameters:} standard deviation $\sigma$, fractional bits $l$, truncation length $\kappa$, failure bound $\delta$, with $\sigma_{\sf int}=2^{l}\sigma$, proposal scale $t=\lfloor\sigma_{\sf int}\rfloor+1$, trial count $M$, and $D=2\sigma_{\sf int}^{2}t^{2}$
\Statex \hspace{-\algorithmicindent}\textbf{Output:} a secret-shared fixed-point noise $\secret{\eta}\in\mathbb{F}_{n,l}$ distributed as $\Func[Gauss](\sigma,l,\kappa)$
\State \textbf{// $M$ proposals, all drawn in parallel}
\For{$j=1$ to $M$}
    \State $\secret{Y_j}\gets\textsc{DLap}(t)$ \Comment{Algorithm~\ref{alg:dlap} before scaling}
    \State $\secret{X_j}\gets\big(|\secret{Y_j}|\cdot t-\sigma_{\sf int}^{2}\big)^{2}$
    \State $\secret{a_j}\gets\textsc{BerExp}(\secret{X_j})$ \Comment{accept bit}
\EndFor
\State $\secret{Z}\gets$ the first $\secret{Y_j}$ whose $\secret{a_j}$ is $1$ \Comment{oblivious selection}
\State \Return $\secret{\eta}\gets 2^{-l}\secret{Z}$
\Statex
\Statex \hspace{-\algorithmicindent}\textbf{Subroutine} $\textsc{BerExp}(\secret{X})$, returning $\secret{a}$ with $\Pr[a=1]=e^{-X/D}$
\For{each bit $\secret{X_i}$ of $\secret{X}$}
    \State $\secret{b_i}\gets\Func[coin]\big(e^{-2^{i}/D}\big)$
\EndFor
\State \Return $\secret{a}\gets\bigwedge_i\neg\big(\secret{X_i}\wedge\neg\secret{b_i}\big)$
\end{algorithmic}
\end{algorithm}

\begin{theorem}\label{thm:dgauss-secure}
In the $\Func[coin]$-hybrid model, the discrete Gaussian sampler securely realizes $\Func[Gauss]$ against a semi-honest adversary corrupting up to $N-1$ parties.
\end{theorem}

Because acceptance and the arithmetic are again a composition of $\Func[coin]$ calls and local Boolean gates, the simulation argument of Appendix~\ref{app:dlap-proof} applies verbatim. The truncation is part of $\Func[Gauss]$, so the simulator reproduces only the coin outputs, the opened Beaver-triple values, and the public gates, which yields an identical view.

\end{document}